# Astrobiology and the Transformation of Scientific Epistemology

Kristina Šekrst

University of Zagreb

Astrobiology occupies an unusual position within the philosophy of science. Confronted with the *n = 1* problem – having only a single example of life to study – it attempts to investigate life beyond Earth while relying entirely on Earth's biosphere as its reference point, a constraint that creates unique epistemic challenges. Unlike traditional sciences with clear predictive frameworks, astrobiology operates as what we might call a *transient science*: a discipline functioning without foundational certainties, relying predominantly on abductive reasoning, and confronting hypotheses that may remain untestable for decades. It is, in essence, a *science of absence* – of evidence, certainty, and analogy – where progress lies in refining conceptual and experimental tools to recognize unfamiliar forms of life. This positions astrobiology alongside emerging fields like artificial intelligence and cognitive science within a broader transformation of how scientific knowledge is constructed when dealing with phenomena that transcend direct empirical access.

## 25. 1 The Nature of Evidence and Theory-Ladenness

In the philosophy of science and epistemology as the theory of knowledge, the concept of evidence takes the throne. In order for a theory to be a theory, it must at least be susceptible to being supported or *falsified* by *evidence*. By *falsifiability*, Popper (1959) means that a scientific theory must be structured in such a way that it can, in principle, be refuted by some possible observation or experiment. That is, a claim is scientific only if there exists a conceivable basic statement that would contradict it.

First, let us discuss evidence itself before examining what it means to be falsified. Even the standard definition of *knowledge* itself – as *justified true belief* – rests on “justification” as its key term, and Kim (1988) states that “the concept of evidence is inseparable from that of justification”. When one thinks of *evidence*, concrete biological substrates or material objects come to mind. However, evidence itself has been understood in various ways: Russell saw it as mental items of immediate consciousness, Quine as stimulation of sensory receptors, and logical positivists as direct descriptions of observable phenomena, while Popper (1959) emphasized singular statements that could potentially falsify laws, such as “There is a black swan in this location”, which would falsify “All swans are white” (Kelly, 2016). This diversity demonstrates that evidence has never been as straightforward as commonly assumed and has never been restricted to direct evidence alone.

As Thomas Kuhn (1962) emphasized, evidence is never entirely neutral. Observations are *theory-laden*: what we count as evidence depends on our conceptual lenses through which we interpret the world around us. A biosignature, such as methane on Mars, for instance, is understood in the context of a web of prior assumptions about geochemistry, atmospheric chemistry, and biology. In astrobiology, this theory-ladenness is amplified because we have no agreed-upon framework for what life universally is, or how its traces should look beyond Earth. When we claim to detect "anomalies" in an exoplanet's environment, we are projecting Earth-based models of life and planetary processes onto alien worlds. This epistemic filter means that even the most sophisticated spectral data carries a particular interpretive burden: we see what our theories and personal expectations allow us to see. When data is ambiguous, scientists may interpret it to fit their expectations. This theory-ladenness becomes especially problematic given astrobiology's reliance on indirect evidence.

## 25.2 The Epistemological Paradox of Astrobiology

Whereas evidence in science typically involves direct observation or experimental verification, astrobiology must navigate a very different epistemic terrain. Here, evidence is typically *indirect*: we *infer* the possibility of life from chemical patterns or atmospheric signatures rather than potential organisms themselves. In contrast, most areas of modern science deal with more direct evidence: scientists can observe phenomena directly or through established methods with clearer connections between observations and conclusions. This indirectness has led to accusations that astrobiology verges on pseudoscience (Jeancolas et al., 2024). Nevertheless, other disciplines face similar challenges. Paleontology, for example, infers past life from fossil fragments that may turn out to be forgeries (Rowe et al., 2001). Similarly, some areas of contemporary physics – notably string theory and multiverse cosmology – struggle with an evidential foundation that is primarily mathematical rather than observational (Read and Le Bihan, 2021).

Astrobiology’s core notion of indirectness invites the danger of *false positives*, as many candidate biosignatures – such as methane, oxygen, or phosphine – can arise from purely abiotic processes. Foote et al. (2023) argue that we need a more systematic approach: developing theory-based universal biosignatures, moving beyond stepwise scales like NASA’s CoLD (Green et al., 2021) toward probabilistic reasoning, and conducting large-scale experiments to better distinguish between biological and abiotic processes.

Smith and Mathis (2023) argue that the absence of a theory of life fundamentally constrains life detection on exoplanets. Since most current biosignature approaches rely on indirect observations (e.g., atmospheric oxygen or methane) that are not unique to life, without a principled theory distinguishing living from non-living processes, these signals cannot serve as decisive evidence for life. This theoretical gap leaves astrobiology vulnerable to both false positives and false negatives: we may misinterpret abiotic processes as biological, or fail to recognize genuinely alien forms of life that operate outside our current conceptual frameworks.

**25. 3 False Positives and False Negatives**

False positives mark Astrobiology's history: the Allan Hills 84001 meteorite (McKay et al., 1996), Viking's ambiguous life detection results (Levin & Straat, 1976), Mars methane variations (Webster et al., 2015), and the controversial phosphine detection on Venus (Greaves et al., 2020). Each controversy reveals how indirect evidence creates interpretive vulnerabilities. The ALH84001 meteorite initially appeared to contain fossilized Martian microorganisms, with carbonate globules resembling fossil bacteria, organic compounds, and magnetite crystal chains similar to modern magnetotactic bacteria, seemingly telling a coherent story of ancient life. However, this interpretation crumbled under scrutiny as terrestrial contamination and abiotic mineral formation processes proved sufficient to explain all observed features. Analogously, the Viking Labeled Release experiment produced results consistent with metabolic activity, yet the failure of companion instruments to detect organic compounds forced scientists to abandon biological explanations in favor of chemical oxidants.

The recent debate over phosphine on Venus shows how even sophisticated observations can be clouded by interpretive uncertainty. Greaves et al. (2020) detected traces of phosphine – a molecule that, on Earth, signals anaerobic life. Despite employing two independent telescopes and appearing robust, subsequent reanalyses questioned whether the signal represented phosphine at all, suggesting instead sulfur dioxide misidentification or instrumental artifacts. Here we see astrobiology's central epistemic dilemma: the same data can support different conclusions depending on our assumptions about atmospheric chemistry, radiative transfer, or instrument calibration. This exemplifies *underdetermination* in the Duhem-Quine sense: different, even incompatible hypotheses can be made to fit the same evidence by adjusting the surrounding assumptions. Duhem (1954) argued that hypotheses cannot be tested in isolation because they form interconnected webs with auxiliary assumptions, while Quine (1951) extended this to claim that our entire system of beliefs faces experience holistically. In other words, the data alone cannot determine which explanation is correct, as what it supports depends on the larger web of background theories – there is no single 'smoking-gun' observation.

However, the epistemological challenge cuts both ways: if astrobiology risks overinterpreting ambiguous signals as biosignatures, it also risks overlooking genuine evidence of life through *false negatives*. Few scientific fields today devote as much attention to false negatives as to false positives, and astrobiology appears to be a notable exception in this regard. The Viking Labeled Release controversy exemplifies this concern. Gilbert Levin, the experiment's principal investigator, maintained that the metabolic signals indicated genuine life, noting that no chemical explanation has fully accounted for all results (Levin & Straat, 2016). Crucially, since subsequent Mars missions risk planetary contamination with terrestrial microorganisms, the Viking experiments may represent our only opportunity to study pristine Martian samples. This sobering *temporal constraint* of an irreversible, one-time window magnifies the epistemological burden of how we interpret ambiguous evidence – an issue that is rarely confronted so directly in contemporary sciences and modern scientific epistemology.

### 25.4 Beyond Falsifiability

The scientific community's rejection of false positives in astrobiological research also reflects a deeper epistemological bias: our theories about what constitutes habitable environments and recognizable life processes. Similarly, our search strategies may be fundamentally constrained by a certain "conceptual speciesism": the tendency to use human categories as the standard against which all other phenomena are measured (Šekrst, 2024), as a more profound manifestation of the *n=1* problem (Mariscal, 2015). With only one example to work from, we search for water-based chemistry, carbon-based molecules, and atmospheric disequilibria that resemble Earth's biosphere. However, life elsewhere might operate through entirely different chemical pathways or leave signatures we have not yet learned to recognize.

*Falsifiability* reveals its limitations when applied to the core claims of astrobiology. Popper (1959) argued that a scientific theory must be framed so that some possible observation could prove it wrong. However, many hypotheses in astrobiology do not easily fit this mold. Consider the proposition that life exists in Europa's subsurface ocean: while this generates testable predictions about biosignatures or metabolic byproducts, negative results cannot definitively falsify the claim. Life might persist in unexplored regions, at inaccessible depths, or through biochemistries our instruments cannot recognize. The spatial and temporal constraints inherent to planetary exploration create what one might call "unfalsifiability through inaccessibility" rather than unfalsifiability *in principle*: we cannot definitively prove Europa lacks life, not because the claim is inherently untestable, but because we cannot yet access the relevant evidence. This suggests that falsifiability alone cannot serve as the criterion for scientific legitimacy when dealing with phenomena beyond our current empirical reach.

Pollock (1974) draws a helpful distinction between two types of criticism: *rebutting defeaters* provide direct evidence against a claim, while *undercutting defeaters* question whether the evidence supports the conclusion without disproving the claim outright. This distinction is crucial for astrobiology because it reveals that most criticisms do not actually disprove the possibility of extraterrestrial life – they question our methods of detection. Most critiques of astrobiological claims operate as undercutting defeaters, asking whether the evidential relationship between observations and life claims is *sound* rather than providing direct evidence against extraterrestrial life itself. The phosphine on Venus debate illustrates this pattern: critics did not demonstrate that Venus cannot harbor life (a rebutting defeater) but rather contested whether the data genuinely indicated phosphine rather than sulfur dioxide (an undercutting defeater). Similarly, Viking's detractors argued that the Labeled Release results could be explained through abiotic chemistry without proving Mars's sterility.

The scientific status of the field may rest not on strict Popperian falsifiability but on what Carroll (2018) describes as "empirical coherence" within broader theoretical structures. Carroll argues that science routinely operates with indirect evidence and coherence-based justification, advocating for a broader epistemic toolkit including abductive reasoning, Bayesian updating, and empirical coherence across related theories. He uses multiverse cosmology – another field criticized for unfalsifiability – to demonstrate that scientific legitimacy can derive from integration into successful theoretical frameworks rather than strict falsification. This demonstrates that astrobiology can achieve epistemic robustness through coherence and explanatory depth rather than traditional falsification criteria.

### 25.5 Astrobiology as Transient Science

Astrobiology represents what we might call a "transient science" – a discipline that lacks the foundational certainties of established fields. It operates under severe observational constraints, relies heavily on abductive reasoning rather than deductive frameworks, and struggles with hypotheses that may remain untestable for decades.

Namely, we are witnessing a paradigm shift (cf. Kuhn, 1962) in how knowledge is constructed. Whereas conventional sciences operate through observer-dependent methodologies with clear objects of study and human-centric frameworks, astrobiology demands distributed, multi-agent systems capable of navigating epistemic uncertainty while remaining open to radically different forms of life and minds.

This transformation parallels contemporary developments in artificial intelligence research, where similar epistemic puzzles emerge. Both fields struggle with “black box” problems: AI systems produce unexpected capabilities from training data through processes we cannot fully explicate, while astrobiology confronts potential alien “black boxes” in the form of non-terrestrial intelligence manifesting through unexpected signs of life. The underdetermination problem becomes particularly acute in both domains as well. In AI, multiple theoretical frameworks can account for observed machine behaviors, while in astrobiology, diverse abiotic and biotic explanations remain consistent with the same observational data. This curious parallel shows that astrobiology may be pioneering epistemological strategies that will prove essential for understanding other forms of non-human intelligence, whether artificial or extraterrestrial.

The deeper issue lies in “conceptual speciesism”: we investigate animal intelligence by comparing it to human cognition, subject machines to human-designed intelligence tests, and search for extraterrestrial life using Earth-based biochemical templates (Šekrst, 2024). This anthropocentric bias creates systematic blind spots that may prevent us from recognizing genuinely alien forms of intelligence or life, allowing us to have more issues with false negatives. If astrobiology truly aims to study universal phenomena rather than terrestrial extensions, it demands “astrophilosophy”: a philosophical framework capable of addressing fundamental questions about reality, knowledge, and existence without privileging human categories (Šekrst, 2024), along with universal biology (Dick, 2020).

This parallel between astrobiology and AI indicates that both fields may be pioneering epistemological strategies essential for understanding forms of intelligence that transcend direct human comprehension. Nevertheless, astrobiology, alongside artificial intelligence, forces us toward a more radical epistemological transformation. We are building epistemologies where we are *no longer the only interpreters*. Machine learning systems discover patterns in data that escape human recognition. At the same time, the search for extraterrestrial life demands frameworks capable of identifying intelligence that might operate through entirely different cognitive architectures.

The *interdisciplinary* character of these emerging fields signals a broader transformation in scientific practice as well. Cognitive science integrates psychology, neuroscience, computer science, and AI research as it seeks to understand minds beyond the human template: animal cognition, artificial intelligence, and potentially extraterrestrial cognition. Astrobiology similarly draws from various sciences and humanities in its search for life elsewhere. If astrobiology succeeds in discovering intelligent extraterrestrial life, the implications will cascade through virtually every scientific discipline: psychology would need frameworks for alien cognitive architectures, philosophy of mind would require new theories of consciousness, and epistemology itself would face knowledge systems that developed independently of human evolutionary pressures.

This convergence demands *epistemic humilit*y – the recognition that our claims to knowledge are fragile and shaped by contingent conditions (Kidd, 2017). In astrobiology, every mission is a rare and often unrepeatable chance to gather evidence. The Viking experiments on Mars demonstrate this clearly: later missions may have contaminated the planet, leaving the 1976 results as our only window into the untouched Martian chemistry. Such scarcity raises the stakes, forcing us to weigh the risk of false positives against the equally real risk of missing genuine signs of life.

Astrobiology reveals the limits of human knowledge more starkly than most sciences. Its reliance on indirect evidence and interpretive challenges reflects the real difficulty of seeking life that might work in ways we have not considered. This is what makes it important: it forces science to confront its own boundaries. Far from being a weakness, this is astrobiology's key insight – that *uncertainty* can be a *methodological* virtue rather than an obstacle. Astrobiology's central lesson may be that intellectually honest science must remain open to possibilities beyond our current understanding. By confronting the unknown, it demonstrates what science becomes when it accepts that the universe likely exceeds what our present theories and methods can capture.